\documentclass[referee,sn-nature]{sn-jnl}

\usepackage{graphicx}%
\usepackage{multirow}%
\usepackage{amsmath,amssymb,amsfonts}%
\usepackage{amsthm}%
\usepackage{mathrsfs}%
\usepackage[title]{appendix}%
\usepackage{xcolor}%
\usepackage{textcomp}%
\usepackage{manyfoot}%
\usepackage{booktabs}%
\usepackage{algorithm}%
\usepackage{algorithmicx}%
\usepackage{algpseudocode}%
\usepackage{listings}%
\usepackage{bookmark}%
\usepackage{lineno}
\usepackage[T1]{fontenc}
\newcounter{video}
\DeclareRobustCommand{\rvideo}[1]{\refstepcounter{video}\thevideo\label{#1}}

\unnumbered

\newcommand{\aap}{Astron. Astrophys.} 
\newcommand{\apj}{Astrophys. J.} 
\newcommand{\apjl}{Astrophys. J. Lett.} 
\newcommand{\aj}{Astron. J.} 
\newcommand{\icarus}{Icarus} 
\newcommand{\nat}{Nature} 
\newcommand{\areps}{Annu. Rev. Earth Planet. Sci.} 
\newcommand{\epsl}{Earth Planet. Sci. Lett.}    
\newcommand{\gca}{Geochim. Cosmochim. Acta} 
\newcommand{\jgrp}{J. Geophys. Res.: Planets}    
\newcommand{\maps}{Meteorit. Planet. Sci.}
\newcommand{\psj}{Planet. Sci. J.}   
\newcommand{\pnas}{Proc. Natl Acad. Sci. USA}   

\begin{document}

\title[Article Title]{Episodic planetesimal disruptions triggered by dissipation of gas disk}

\author[1,2]{\fnm{Kang} \sur{Shuai}}\email{shuaikang@nju.edu.cn}
\author*[1,2]{\fnm{Li-Yong} \sur{Zhou}}\email{zhouly@nju.edu.cn}
\author[3,4]{\fnm{Hejiu} \sur{Hui}}\email{hhui@nju.edu.cn}

\affil*[1]{\orgdiv{School of Astronomy and Space Science}, \orgname{Nanjing University}, \orgaddress{\street{163 Xianlin Road}, \city{Nanjing}, \postcode{210023}, \country{China}}}
\affil[2]{\orgdiv{Key Laboratory of Modern Astronomy and Astrophysics in Ministry of Education}, \orgname{Nanjing University}, \orgaddress{\street{163 Xianlin Road}, \city{Nanjing}, \postcode{210023}, \country{China}}}
\affil[3]{\orgdiv{State Key Laboratory of Critical Earth Material Cycling and Mineral Deposits \& Lunar and Planetary Science Institute, School of Earth Sciences and Engineering}, \orgname{Nanjing University}, \orgaddress{\street{163 Xianlin Road}, \city{Nanjing}, \postcode{210023}, \country{China}}}
\affil[4]{\orgname{CAS Center for Excellence in Comparative Planetology}, \orgaddress{\street{96 Jinzhai Road}, \city{Hefei}, \postcode{230026}, \country{China}}}

\abstract{Catastrophic disruptions of planetesimals occur in high-velocity collisions. Radioisotope dating of planetesimal disruption events recorded in meteorites confirms frequent catastrophic collisions in the first 10~Myr of the Solar System, reflecting a violent environment of the time. However, the nebula gas can damp the eccentricity of planetesimals and suppress the frequency of planetesimal collisions. Strong dynamical mechanisms that excited the protoplanetary disk are required. Here we show that the sweeping secular resonances of Jupiter and Saturn induced by the nebular gas dissipation, together with the mean motion resonances of Jupiter, can trigger a large number of catastrophic collisions, which occur episodically when the secular resonances are at 2--3 astronomical units and continue thereafter. After the gas dissipation completes, catastrophic collisions decrease in frequency, with scattering by planetary embryos becoming the major driving force of the collisions. Our results suggest that the violent environment excited by secular and mean motion resonances can be ubiquitous in protoplanetary disks during nebula dissipation.}


\maketitle

\section{Introduction}\label{sec1}
Collisions between rocky bodies shape the orbital architecture, composition, and internal structure of terrestrial planets and asteroids~\citep{Asphaug2010, Gabriel2023}. High-velocity collisions can cause catastrophic disruptions of rocky bodies, leading to mantle and crust stripping, fragment ejection, and vaporization~\citep{Marcus2009, Genda2017, Carter2018, Davies2020, Allibert2021}. The first 10--20~Myr of the Solar System, when both the frequency and energy of collisions are higher than other time, as supported by the radioisotope dating of collision events, is the most important period for collisions~\citep{Davison2013}. Chondrules and metal grains in CB chondrites are supposed to form in high-velocity collisions capable of vaporizing metal from a planetesimal’s core~\citep{Krot2005}. These collisions are inferred to have occurred around 4~Myr after the Solar System formation, based on the formation age of CB chondrites~\citep{Krot2005, Bollard2015, Wolfer2023, Pravdivtseva2017, Yamashita2010}. Moreover, the times of catastrophic collisions that stripped the mantles of the parent bodies of iron meteorites, have be determined based on the closure times of $^{107}$Pd--$^{107}$Ag decay system, indicating an energetic inner Solar System during approximately 7.8--11.7~Myr after the Solar System formation~\citep{Hunt2022}.
 
However, planetesimal dynamics in the first 10--20~Myr of the Solar System is subject to various processes, and their combined effects are under debate. Particularly, the nebula gas damps the orbital eccentricity of rocky bodies and thus stabilizes their orbits, suppressing the impact velocity of collisions~\citep{Zhou2007, Walsh2019}. Therefore, some strong dynamical mechanisms are required to counter the damping effects and make the high-speed collisions possible. Generally, the gravitational influence of the giant planets may excite the orbits of rocky bodies~\citep{Johnson2016, Oshino2019, Carter2020}. The sweeping secular resonances were proposed to pump up the orbital eccentricity of planetesimals and lead to the mass depletion of the main asteroid belt~\citep{Lecar1997, Nagasawa2005, Zheng2017}. In planet formation simulations with and without a gas disk, the dynamics of planetesimals is significantly influenced by the sweeping secular resonances and mean motion resonances (MMRs), respectively~\citep{Woo2021}. In addition, the growing planetary embryos may also exert dynamical excitation on planetesimals and other embryos~\citep{Ida2010, O'Brien2007}. However, the planetesimal collisions under the combined effects of these mechanisms have not been investigated specifically.

Here we consider orbital evolution of planetesimals in a dissipating gas disk under the gravitations of Jupiter and Saturn. The correlations between collisional events of different energies, the planetesimals' orbital evolution, and the locations of secular and MMRs found in our simulations suggest that these resonances are the dominant driving forces of the planetesimal disruptions in the first 10~Myr of the Solar System.

\section{Results and discussion}\label{sec2}
In a protoplanetary disk with nebula gas, the gas disk exerts perturbations on both planets and planetesimals, via the gravitational force and the damping effects~\citep{Lecar1997, Nagasawa2005, Zheng2017}. The strength of these perturbations decreases with time due to gas dissipation. We performed high-resolution N-body simulations that included the effects of the gas disk and directly resolved planetesimal-planetesimal interactions (see Methods, subsection N-body simulations). The initial conditions include equal-mass planetesimals along with Jupiter and Saturn. In our nominal simulations, 73,982 initial planetesimals with radius of 280~km are used, Jupiter and Saturn are initially placed on their current orbits, and the gas depletes uniformly throughout the disk. We also conducted test simulations considering different numbers and sizes of planetesimals, giant planet configurations, gas depletion models, and the inclusion of initial planetary embryos. The orbits of planetesimals were tracked and the catastrophic collisions were monitored. In our simulations, Jupiter and Saturn remain stable on orbits that resemble closely their initial orbits, while the planetesimals' orbits are strongly affected by the sweeping secular apsidal resonances and the gas damping effects. From a quiet initial disk of planetesimals, the system is excited and high-velocity collisions develop.

\begin{figure}
\centering
\includegraphics[width=0.6\textwidth]{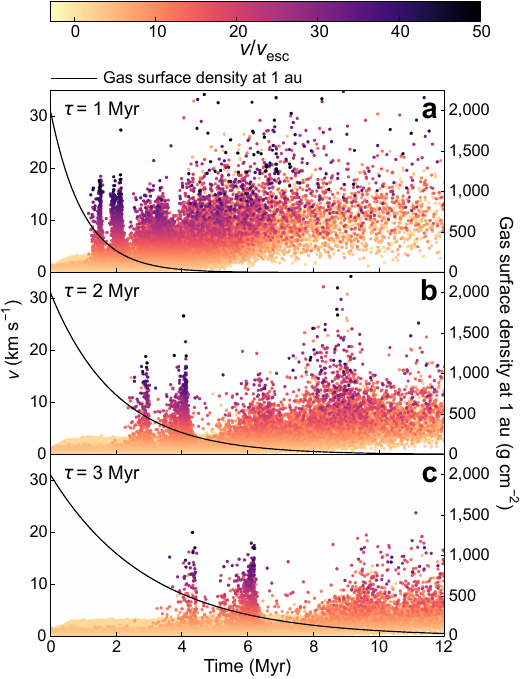}
\caption{Temporal evolution of impact velocity and the surface density of the gas disk. Three simulations with gas dissipation timescale ($\tau$) of 1~Myr (\textbf{a}), 2~Myr (\textbf{b}), and 3~Myr (\textbf{c}) are shown. Each dot is one collision event, with colors representing the ratio of impact velocity to the mutual escape velocity of impactors ($v/v_{\rm esc}$). The solid curves show the gas surface density at 1~au. The solid and open triangles on the $x$-axes represent 1.2$\tau$ and 2.4$\tau$, which are the beginning of episodic (Episode I) and continuous (Episode II) high-velocity collisions, respectively. Source data are provided as a Source Data file.
}
\label{figure1}
\end{figure}

\subsection{Evolution of impact velocity}\label{subsec2-1}

More than 60,000 collisions among planetesimals were detected in each simulation, enabling a statistical analysis on the collision frequency, impact velocity, and orbital evolution of impactors. The temporal evolution of impact velocity ($v$) correlates with the timescale of nebula dissipation ($\tau$), with more delayed high-velocity collisions for larger $\tau$ (slower nebula dissipation) (Fig.~\ref{figure1}). In the first 1.2$\tau$, only low-velocity ($\lesssim$2.5~km\,s$^{-1}$) collisions occur. Between $1.2-2.4\tau$, multiple short time intervals with frequent high-velocity collisions form spikes of impact velocity. High-velocity collisions are rare between these spikes. After 2.4$\tau$, high-velocity collisions occur continuously, while low-velocity collisions become rare after about $5\tau$, along with the whole system being excited. On the other hand, planetary embryos (defined as masses greater than 0.01 Earth masses) form within the first 2--3~Myr in all simulations. Following this runaway growth phase, the embryos grow slower until gas dissipation completes, when giant collisions between embryos begin. Although planetary embryos form on similar timescales, the high-velocity collisions show varying and $\tau$-dependent temporal evolution, suggesting that embryo scattering is not the dominate driver of such events during gas dissipation.

Given the timescale $\tau$ of gas dissipation, the correlation between collision occurrence and $\tau$ indicates that these high-velocity collisions happen when the gas surface density reaches specific levels (Fig.~\ref{figure1}). The secular resonances, occurring when the apsidal precession rate of a planetesimal matches that of a giant planet, sweep inward across the inner region of Solar System during nebula dissipation~\citep{Heppenheimer1980, Ward1981, Lecar1997, Nagasawa2000}. The inward sweeping of secular resonances is driven by changing gravitational potential of the nebula gas with decreasing gas density, thus has the same timescale as nebula dissipation. Therefore, the correlation between collisions and $\tau$ as shown in Fig.~\ref{figure1} suggests that the high-velocity collisions might be stimulated by the sweeping secular resonances.

\begin{figure}
\centering
\includegraphics[width=\textwidth]{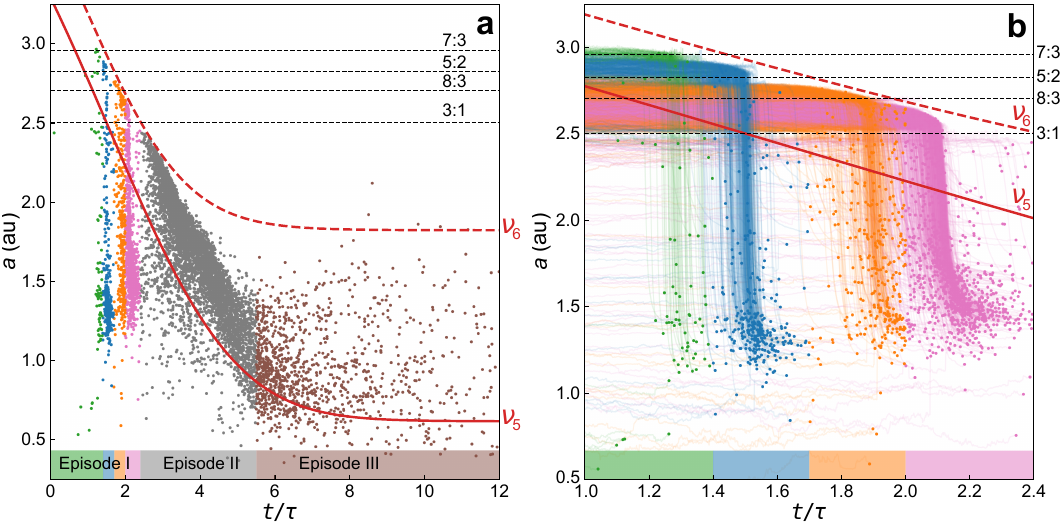}
\caption{Semi-major axes of projectiles in catastrophic collisions. The $x$-axis is the ratio of time to gas dissipation timescale ($t/\tau$). The results of the simulation with $\tau=1$~Myr are shown. The sweeping secular resonances ($\nu_5$ and $\nu_6$, nominal locations shown in solid and dashed red curves) and four Jovian MMRs (dashed black lines) are plotted. Each dot represents one collision, color-coded based on the collision time, as indicated by the discrete color bar on the $x$-axis. (\textbf{a}) Instantaneous semi-major axes of projectiles in catastrophic collisions at the time of collision. The catastrophic collisions are divided into Episode I ($t\leq2.4\tau$), Episode II ($2.4\tau<t\leq5.5\tau$), and Episode III ($t>5.5\tau$). (\textbf{b}) Evolution of semi-major axes of projectiles in Episode I collisions. The curves show the evolution of semi-major axes, ending by dots indicating the collisions. Source data are provided as a Source Data file.
}
\label{figure2}
\end{figure}

\subsection{Collision episodes and driving forces}\label{subsec2-2}

Based on the impact velocity, impact angle, and impactor's size, the collisions can be classified using an analytical model~\citep{Leinhardt2012}. Those that have reduced mass kinetic energy larger than the critical energy for catastrophic disruption are defined as catastrophic collisions~\citep{Stewart2009}. For each catastrophic collision, we define one of the two impactors that has a larger eccentricity right before the collision as the projectile, and the other one as the target. Generally, the projectile has an excited orbit and collides with the target that has a more stable orbit. Thus we can attribute the collision to the dynamical mechanism that excites the projectile's orbit. For most catastrophic collisions, the projectiles are those planetesimals that have been just excited by the sweeping secular apsidal resonance of Jupiter ($\nu_5$), as the correlation between their semi-major axes and the nominal location of $\nu_5$ (Fig.~\ref{figure2}) reveals. Note that the semi-major axes of projectiles at collision are not necessarily equal to the locations (distance to the Sun) where collisions occur, but the collision locations show a similar spatial and temporal distribution, which also correlates with $\nu_5$ (Supplementary Fig.~8).

The catastrophic collisions can be divided into three major episodes. For the catastrophic collisions that occur before 2.4$\tau$ (hereafter Episode I), most of the projectiles migrate inward rapidly across the $\nu_5$ resonance, and collide with the target (Fig.~\ref{figure2}b and Supplementary Movie~\ref{v1}). This inward migration has also been observed in simulations using massless planetesimals~\citep{Lecar1997, Nagasawa2005, Zheng2017, Gong2019}. The Episode I collisions occur episodically, corresponding to the spikes in Fig.~\ref{figure1}. However, the catastrophic collisions between $2.4-5.5\tau$ (Episode II) occur continuously. Most Episode II collisions occur near the $\nu_5$ resonance (Supplementary Fig.~8), and the semi-major axes of projectiles at collision are within 1 astronomical unit (au) from $\nu_5$ (Fig.~\ref{figure2}a). Therefore, the Episode I and II collisions are mainly driven by the $\nu_5$ resonance. The Episode III collisions occur after 5.5$\tau$, when the nebula gas has mostly dissipated and the secular resonances nearly stop migrating. These collisions are spatially dispersed, independent of the location of $\nu_5$ (Fig.~\ref{figure2}a). In this period, planetary embryos have formed and their orbits are excited by mutual gravitational perturbations~\citep{Chambers1998}. The Episode III collisions are likely driven by scattering of planetesimals by planetary embryos, as evidenced by the scarcity of low-velocity collisions in this period (Fig.~\ref{figure1}).

The episodic occurrence of Episode I collisions suggests that they are not only driven by the secular resonances, but also affected by other dynamical mechanics. The collision times are found to be related to the initial semi-major axes of projectiles. The projectiles in the earlier collisions have larger initial semi-major axes (Fig.~\ref{figure2}b), with only exceptions of collisions at about 1.9$\tau$ with initial semi-major axes a little larger than 2.5~au (the 3:1 resonance with Jupiter). Orbital evolution of projectiles (Fig.~\ref{figure2}b) show that the semi-major axes of projectiles slowly decrease until crossing an MMR, after which the projectiles migrate inward rapidly and cause frequent catastrophic collisions in short time intervals. When collisions aroused by an MMR ceased, the catastrophic collisions are very rare, until projectiles with smaller initial semi-major axes get close to another MMR. These temporal gaps of catastrophic collisions correspond to the gaps of semi-major axes of planetesimals around the Jovian MMRs (Supplementary Fig.~9).

\begin{figure}
\centering
\includegraphics[width=\textwidth]{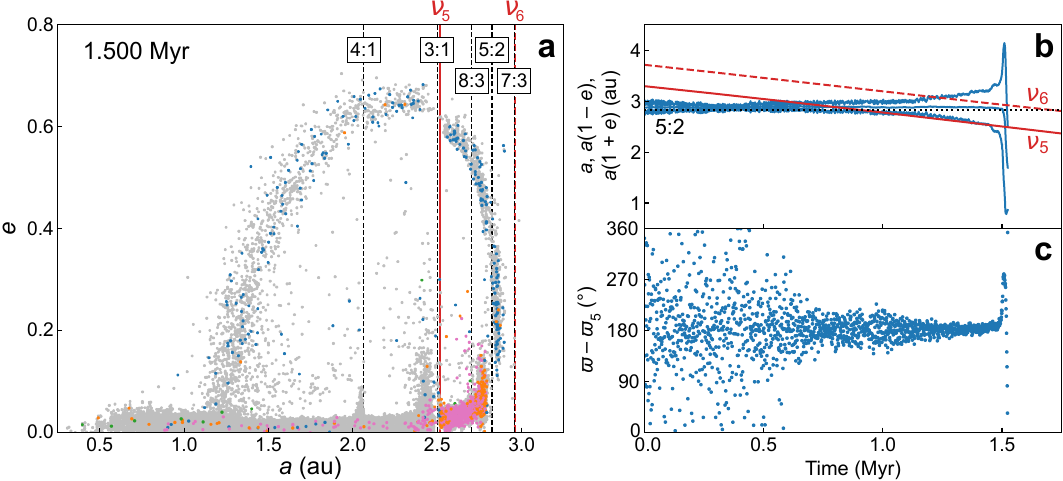}
\caption{Orbital evolution of planetesimals in Episode I. Shown are the results of a simulation with $\tau=1$~Myr, including semi-major axis ($a$), eccentricity ($e$), and the difference in longitudes of periapsis between a planetesimal and Jupiter ($\varpi-\varpi_5$). The solid and dashed red lines show the locations of $\nu_5$ and $\nu_6$. The black lines are Jovian MMRs. (\textbf{a}) A snapshot of semi-major axis and eccentricity evolution of planetesimals at 1.5~Myr (Supplementary Movie~\ref{v2}). Each dot represents a planetesimal. The projectiles in Episode I catastrophic collisions are color-coded as in Fig.~\ref{figure2}. Other planetesimals are gray. (\textbf{b}) Orbital evolution of a projectile of catastrophic collision. The projectile gains eccentricity exceeding 0.6 at 5:2 MMR of Jupiter and collides with a planetesimal at 1.52~Myr. (\textbf{c}) Evolution of the critical angle of secular resonance $\nu_5$ for the projectile. Source data are provided as a Source Data file.
}
\label{figure3}
\end{figure}

In fact, not only those projectiles of catastrophic collisions, but all planetesimals with initial semi-major axes larger than 2\,au migrate inward (Fig.~\ref{figure3}a and Supplementary Movie~\ref{v2}). This clearing process by sweeping secular resonances and gas damping agree with the results of previous studies~\citep{Lecar1997, Nagasawa2005, Zheng2017, Woo2021}, but the induced collisions and the simultaneous effects of MMRs received less attention. The difference in longitudes of periapsis between the projectiles and Jupiter ($\varpi-\varpi_5$) shows that the projectiles are locked in the $\nu_5$ resonance before the inward migration (Fig.~\ref{figure3}c). The secular resonance with Saturn ($\nu_6$) is less effective than $\nu_5$ since such locking in $\nu_6$ has not been observed. Trapped in $\nu_5$, the planetesimals' eccentricities grow slowly with oscillation (Fig.~\ref{figure3}b). With eccentricities of approximately 0.1, the planetesimals migrate slowly due to the damping effect, approaching a Jovian MMR. For planetesimals with moderate eccentricity, the widths of Jovian inner MMRs increase with growing eccentricity~\citep{Murray1999}. These planetesimals are easily captured by an MMR, in which their eccentricities increase rapidly to values exceeding 0.6 (Fig.~\ref{figure3}a and b). Therefore, the interplay of the effects of $\nu_5$, MMRs, and damping on planetesimals enhances the eccentricity excitation. Then, the semi-major axes and eccentricities drop rapidly by gas damping, and the projectiles escape from $\nu_5$ (Fig.~\ref{figure3}c and Supplementary Movie~\ref{v3}). Shortly after this rapid inward migration, the catastrophic collisions occur (Supplementary Movie~\ref{v1}). Therefore, the combined effects of secular and mean motion resonances trigger the episodic catastrophic collisions in Episode I, leading to the spikes of high impact velocities in Fig.~\ref{figure1}.

\begin{figure}
\centering
\includegraphics[width=0.6\textwidth]{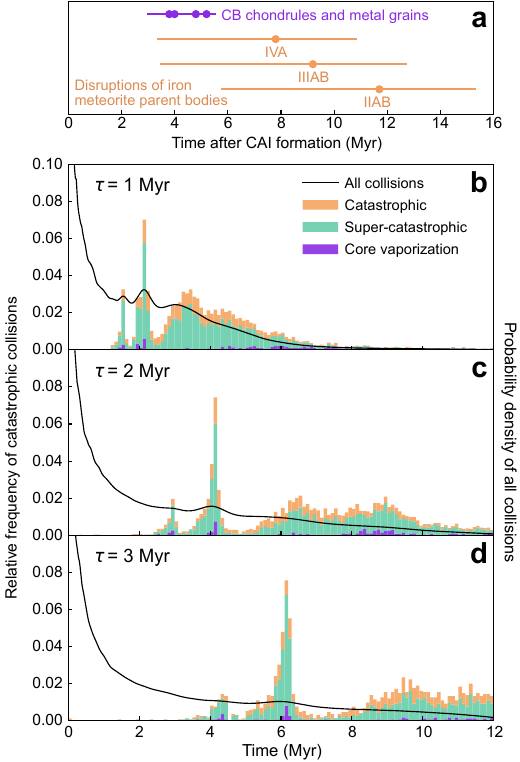}
\caption{Comparison between the radioisotope ages of collision events and the simulated frequency of catastrophic collisions. (\textbf{a}) Radioisotope ages of collision events. The ages of CB chondrules and metal grains dated using different isotope systems~\citep{Bollard2015, Wolfer2023, Pravdivtseva2017, Yamashita2010} and the disruption times of three iron meteorite parent bodies~\citep{Hunt2022} are shown. (\textbf{b-d}) Temporal evolution of collision frequency. Three simulations using different gas dissipation timescales ($\tau$) are shown. Every 0.1~Myr, the frequencies of catastrophic collisions, super-catastrophic collisions, and super-catastrophic collisions with impact velocity $>$13~km\,s$^{-1}$ that potentially cause vaporization of planetesimal cores are shown relative to the total number of catastrophic collisions. The kernel density estimates of all collisions (black curves) are plotted with a different scale. Source data are provided as a Source Data file.
}
\label{figure4}
\end{figure}

\subsection{Frequency of catastrophic collisions}\label{subsec2-3}
Contrary to the exponentially decline of the overall collision frequency, the occurring frequency of catastrophic collisions varies in a non-monotonic way (Fig.~\ref{figure4}). The catastrophic collisions happen most frequently at times around $2\tau$ when $\nu_5$ and $\nu_6$ cross the MMRs in Episode I and remain frequent before the complete gas dissipation in Episode II. In Episode III, the catastrophic collisions decrease with planetesimals being cleared out by grown planetary embryos. In the most energetic collisions, the mass of the largest remnant is smaller than 1/10 of the total mass of impactors, and these collisions are super-catastrophic collisions~\citep{Leinhardt2012}. The fraction of super-catastrophic collisions in all collisions reaches 0.2 or greater in most of time (Supplementary Fig.~10). If the gravitational potential of the gas disk is not included in the model, and thus the secular resonances do not sweep inward, the fraction of super-catastrophic collisions is much lower (Supplementary information section 4). Therefore, the inclusion of gravitational potential in our model results in higher maximum impact velocities and more frequent planetesimal disruptions than previous simulations that do not include giant planet migration~\citep{Carter2015, Johnson2016}. On the other hand, in the Grand Tack model, Jupiter migrates inward and then outward, strongly exciting planetesimal orbits and triggering high-velocity collisions~\citep{Walsh2011, Carter2015, Johnson2016}. The fraction of super-catastrophic collisions under the Grand Tack model is about 0.2~\citep{Carter2015}, similar to our results. Recent hydrodynamic simulations show that Jupiter and Saturn become locked in 2:1 MMR during inward migration in a low-viscosity disk, contradicting the Grand Tack model, which requires them to be captured in the 3:2 MMR~\citep{Griveaud2024}. Our results here suggest that the protoplanetary disk without giant planet migration could be more violent than previously thought, comparable to the disk excited by the migration of Jupiter under the Grand Tack model.

After 5$\tau$, the collision frequency deceases to a low level (Fig.~\ref{figure4}), albeit the fraction of catastrophic collisions remains high (Supplementary Fig.~10). Therefore, the most violent period is the first $1-5\tau$. Similar to the catastrophic collisions, the catastrophic hit-and-run events have non-monotonic frequency as well (Supplementary Fig.~11). These collisions are less energetic than catastrophic ones but have large impact angles, which potentially disrupt the smaller one of the two impactors~\citep{Leinhardt2012}.

\subsection{Meteorite records of collisions}\label{subsec2-4}
Comparing the results of simulated collisions with the meteorite records of planetesimal disruptions provides insights into the evolution of the protoplanetary disk. CB chondrites, one of the youngest dated chondrite groups, are mainly composed of non-porphyritic chondrules and zoned Fe-Ni metal grains~\citep{Krot2005}. These components of CB chondrites formed from the vapor-melt plume in a collision that vaporized the metal core~\citep{Fedkin2015}, which requires an impact velocity of at least 18$\pm$5~km\,s$^{-1}$~\citep{Kraus2015}. This threshold would be lower in the ambient pressure of solar nebula~\citep{Davies2020}. Therefore, it is reasonable to use the lower bound of this range (13~km\,s$^{-1}$). Although the source region of CB chondrites is unclear, the collision potentially occurred in the main asteroid belt~\citep{Johnson2016}. If the colliding planetesimals originated from the outer Solar System, the growth of giant planets could have scattered them into the main belt~\citep{Raymond2017, Nesvorny2024}. In addition, the accretion of post-impact material into the CB parent body requires incomplete gas dissipation~\citep{Johnson2016}. The discovery of chondrules of nebular origin in CB chondrites further supports that CB chondrites formed before complete gas dissipation~\citep{Mahle2024}. Radioisotope dating shows that CB chondrules and metal grains formed at about 4~Myr after the Solar System formation~\citep{Krot2005, Bollard2015, Wolfer2023, Pravdivtseva2017, Yamashita2010}. On the other hand, the disruptions of iron meteorite parent bodies occurred later during approximately 7.8--11.7~Myr after the Solar System formation, as determined using the $^{107}$Pd--$^{107}$Ag decay system~\citep{Hunt2022}. Our modeled catastrophic collisions in the first 12~Myr of the Solar System might be a reasonable birth place for CB chondrites and the cause of planetesimal disruptions. In our simulations, the asteroid belt is depleted by the sweeping secular resonances and the collisional remnants may reside in the terrestrial planet region (Supplementary Movie 2). A fraction of these remnants could have been scattered into the asteroid belt, serving as the source asteroids of meteorites. This asteroidal implantation does not require the gas disk and could have occurred after gas dissipation~\citep{Bottke2006, Raymond2017}.
 
We selected the super-catastrophic collisions with impact velocities exceeding 13~km\,s$^{-1}$, which may cause vaporization of planetesimal cores and potentially formed the CB chondrites (Fig.~\ref{figure4}). These collisions mainly occur in Episode I or later than 4$\tau$. However, the frequency of catastrophic collisions declines after 4.5$\tau$. Considering the radioisotope ages of collision events and the required residual nebula gas for CB parent body accretion, CB chondrites could have formed in an Episode I collision, and the Episodes II and III collisions could have resulted in the disruptions of iron meteorite parent bodies. The modeled collision frequency in simulations with $\tau=2$~Myr appears to match the meteorite records (Fig.~\ref{figure4}). However, the time zero in our simulations could be different from the time of calcium- and aluminum-rich inclusion (CAI) formation, which is considered the time zero of the Solar System. In addition, the start time of gas dissipation varies in different models~\citep{Carter2015}. Nevertheless, our results show strong correlation between the sweeping secular resonances and catastrophic collisions. The catastrophic collisions occur episodically when the secular resonances cross the Jovian MMRs, and occur continuously when the secular resonances approach their current location. These conclusions are independent of the time zero and the start time of gas dissipation.

Our results demonstrate that the protoplanetary disk without giant planet migration is not as calm as previously thought. High-velocity collisions occur frequently in the first 10~Myr of the Solar System regardless of whether the giant planets have stable orbits. Our results do not exclude other potential causes of high-velocity collisions. Under the Grand Tack scenario~\citep{Walsh2011}, the migration of Jupiter could result in high-velocity collisions, including those capable of vaporizing the planetesimal cores and potentially forming CB chondrites~\citep{Johnson2016, Carter2020}. The giant planet instability potentially led to the disruptions of iron meteorite parent bodies~\citep{Hunt2022}, but when this instability occurred is under debate~\citep{Liu2022a, Avdellidou2024}. Nevertheless, our results suggest that the combined effects of sweeping secular resonances and MMRs can excite the planetesimal orbits and cause the catastrophic collisions, which can be ubiquitous in protoplanetary disks during gas dissipation. In addition to forming meteorite components and disrupting planetesimals, high-velocity collisions can cause melting and vaporization~\citep{Kraus2015, Davies2020}. Planetesimals in the Solar System may have experienced substantial collision-induced vapor loss, as evidenced by isotopic signatures in meteorites~\citep{Hin2017, Koefoed2022, Zhu2025}. Our results provide the dynamical mechanism that triggered the high-velocity collisions recorded in these isotopic signatures.

\section{Methods}\label{sec3}
\subsection{N-body simulations}\label{sec3-1}
To simulate planetesimal collisions in the first 12~Myr of the Solar System, we performed high-resolution N-body simulations using a GPU-accelerated code GENGA~\citep{Grimm2014, Grimm2022}. In most previous studies of planetesimal evolution under the effects of sweeping secular resonances, massless particles were used to represent planetesimals, and planetesimal collisions were neglected~\citep{Nagasawa2005, Zheng2017, Gong2019}. Most previous studies that simulated collisional evolution of planetesimals used statistical methods to model the collisions~\citep{Johnson2016, Walsh2019}, or used artificial expansion of planetesimal radii to increase the collision frequency and speed up the calculation~\citep{Kokubo1998, Carter2015}. In contrast to these methods, we adopted the full interactive gravity mode of GENGA, with all bodies feeling the gravity of each other. A large number of close encounters were simulated using the Bulirsch--Stoer method, with collisions identified directly based on the radii and separations of the encountering bodies~\citep{Grimm2014}. The direct computation of planetesimal-planetesimal interactions requires much more computation resources than neglecting them~\citep{Raymond2006}, but makes it possible to simulate planetesimal-planetesimal collisions accurately~\citep{Oshino2019, Carter2020}.

At the beginning of each simulation, 73,982 planetesimals with radius of 280~km were used. The solid surface density profile follows the minimum-mass solar nebula (MMSN) model~\citep{Hayashi1981}:
\begin{linenomath}
    \begin{equation}
        \label{eq:1}
        \Sigma(r)=\Sigma_1\left(\frac{r}{1\rm{\, au}}\right)^{-3/2},
    \end{equation}
\end{linenomath}
where $\Sigma_1=7$ g\,cm$^{-2}$ is the solid surface density at 1~au. The total mass of planetesimals is 3.4~$M_{\rm Earth}$, which initially extends from 0.5~au to 3~au. The initial orbits of planetesimals are nearly circular (eccentricity $<$~0.02) and coplanar (inclination $<$ 1$^\circ$). The other initial angular orbital elements of planetesimals (longitude of the ascending node, argument of periapsis, and mean anomaly) were chosen uniformly at random from 0$^\circ$ to 360$^\circ$. In addition to planetesimals, Jupiter and Saturn were included, which initially resided on their current orbits with semi-major axes of 5.20~au and 9.58~au, eccentricities of 0.049 and 0.056, and inclinations of 0.33$^\circ$ and 0.93$^\circ$ with respect to the invariant plane.

We conducted test simulations to check if high-velocity collisions occur with different configurations of giant planets (Supplementary information section 2). Recent studies suggest that Jupiter and Saturn likely locked in the mutual 2:1 MMR in a low-viscosity disk~\citep{Griveaud2024}. This resonant configuration would represent a more physically realistic state for the protoplanetary disk than the present-day orbits we used in our nominal simulations. Our test simulations with Jupiter and Saturn in the 2:1 resonance (Supplementary Fig. 2) show episodic high-velocity collisions similar to simulations with current giant planet orbits, consistent with a recent study suggesting that sweeping secular resonances occur in such giant planet configuration~\citep{Goldberg2026}. Similar results have been obtained in the test simulation using initially circular orbits for Jupiter and Saturn (Supplementary Fig. 3). Our conclusions from the simulations using the current giant planet orbits remain valid in these configurations.
 
A large number of close encounters in the simulations are very computationally expensive. Intermediate levels and sub-steps were used to shorten time steps for the symplectic integration, which reduced the number of bodies in each close-encounter group, mitigating the computation cost of the Bulirsch--Stoer method~\citep{Grimm2022}. At the beginning of each simulation, three intermediate levels and four sub-steps were used. These parameters were adjusted to ensure that no close-encounter group contained more than 8--16 bodies, a configuration that our pretests indicate yields optimal performance on an NVIDIA V100 GPU. Nevertheless, each high-resolution run (starting with 73,982 planetesimals) requires over four months of continuous computation. In addition to the high-resolution simulations, we performed test simulations to study the influences of number and size of planetesimals and the inclusion of initial planetary embryos (Supplementary information section 1). The gas drag effect varies with planetesimal size, and embryos exert perturbation on planetesimals from the start of the simulations. Nevertheless, the collision episodes and their driving forces remain similar with those in the high-resolution simulations. For the simulations with initial embryos, we conducted an additional test with the planetesimal-planetesimal interactions disabled. The lower maximum impact velocity and frequency of high-velocity collisions confirm that the planetesimal-planetesimal interactions are necessary to reproduce the planetesimal disruption events recorded in meteorites (Supplementary information section 4).

The gas disk model implemented in GENGA was used, which follows ref. \cite{Morishima2010}. The effects of gas disk include the gravitational potential, hydrodynamic drag, and planet-disk tidal interaction~\citep{Morishima2010}. All planetesimals and giant planets feel the gravitational potential of the gas disk. The hydrodynamic drag and planet-disk tidal interaction damp the orbital eccentricities and inclinations of planetesimals. We do not use the gas drag enhancement~\citep{Morishima2010} in our simulations to ensure accurate dynamical evolution of planetesimals.

The gas disk profile follows a power law, and the gas surface density decreases exponentially in time and uniformly throughout the disk:
\begin{linenomath}
    \begin{equation}
        \label{eq:2}
        \Sigma_{\rm gas}(r,t)=\Sigma_{\rm gas,1}\left(\frac{r}{1\rm{\, au}}\right)^{-p}\exp\left(-\frac{t}{\tau}\right),
    \end{equation}
\end{linenomath}
where $t$ is time, $\Sigma_{\rm gas,1}$ is the initial gas surface density at 1~au, and $\tau$ is the timescale of gas dissipation. In the three high-resolution simulations, $\tau$ was set as 1~Myr, 2~Myr, and 3~Myr. $\Sigma_{\rm gas,1}$ is assumed to be 2,000 g\,cm$^{-2}$ and $p=1$~\citep{Morishima2010}, different from the $\Sigma_{\rm gas,1}$ value (1700 g\,cm$^{-2}$) and $p=1.5$ in the MMSN model~\citep{Hayashi1981}. This gas density profile is adopted for comparison with previous simulations that included a gas disk~\citep{Morishima2010, Woo2021}. The scale height is given by $h=c/\Omega_{\rm kep}$, where $c$ is isothermal sound velocity and $\Omega_{\rm kep}$ is Keplerian frequency. At 1~au, $h=0.03358$~au for $c=1$~km\,s$^{-1}$. The gas density is given by
\begin{linenomath}
    \begin{equation}
        \label{eq:2_1}
        \rho_{\rm gas}\left(r,z\right)=\frac{\Sigma_{\rm gas}}{\sqrt{2\pi}h}\exp\left({-\frac{z^2}{2h^2}}\right),
    \end{equation}
\end{linenomath}
where $z$ is the vertical distance from the midplane of the protoplanetary disk. In addition to this uniform depletion model, we performed test simulations to check whether our conclusions are valid if the gas disk did not dissipate uniformly. The gap-opening model with an extending disk gap centered at Jupiter and the inside-out model with an outward moving inner edge of the gas disk~\citep{Nagasawa2000} are used. High-velocity collisions occur in both tests and the collision times correlate with the extending rate of disk gap and the moving rate of the disk inner edge (Supplementary information section 3). These correlations are similar to the correlation between collision times and $\tau$ under the uniform depletion model (Fig.~\ref{figure1}). Therefore, different gas dissipation models do not substantially alter the occurrence of high-velocity collisions and their driving forces.

\subsection{Classification of catastrophic collisions}\label{sec3-2}
The simulated collisions can be classified using the EDACM collision model~\citep{Leinhardt2012}. The potential collision outcomes were calculated based on the impact velocity ($v$), impact angle ($\theta$), and the masses of the larger impactor ($M_{\rm l}$) and the smaller impactor ($M_{\rm s}$). Although these outcomes were not used in the simulations, they are useful to classify the collisions into different regimes. The mass of the largest remnant $M_{\rm lr}$ can be calculated using the universal law generalized to all impact angles~\citep{Leinhardt2012}:
\begin{linenomath}
    \begin{equation}
        \label{eq:3}
        \frac{M_{\rm lr}}{M_{\rm l}+M_{\rm s}}=-0.5\left(\frac{Q_{\rm R}}{Q^{\prime*}_{\rm RD}}-1\right)+0.5,
    \end{equation}
\end{linenomath}
where $Q_{\rm R}=0.5v^2M_{\rm l}M_{\rm s}/(M_{\rm l}+M_{\rm s})^2$ is the reduced mass kinetic energy and $Q^{\prime*}_{\rm RD}$ is the critical energy for catastrophic disruption. $Q^{\prime*}_{\rm RD}$ can be calculated using the masses and radii of impactors, impact angle, and material parameters $c^*$ and $\bar\mu$~\citep{Leinhardt2012}. The energy dissipation parameter $c^*$ and the coupling parameter $\bar\mu$ characterize the physical properties of impactors and vary with impactor size. If the spherical radius of the total impactor mass at a density of 1 g\,cm$^{-3}$ ($R_{\rm C1}$) is smaller than 1,000~km, $c^*=5.0$ and $\bar\mu=0.37$ are used; for $R_{\rm C1}\geq1,000$~km, $c^*=1.9$ and $\bar\mu=0.36$ are used.

The collisions with the impact parameter $b$ ($b={\rm sin}\theta$) larger than the critical impact parameter $b_{\rm crit}$ ($b_{\rm crit}=R_{\rm l}/(R_{\rm l}+R_{\rm s})$, where $R_{\rm l}$ and $R_{\rm s}$ are the radii of the larger and the smaller impactors~\citep{Asphaug2010}) and $M_{\rm lr}\geq M_{\rm l}$ are classified within the hit-and-run regime~\citep{Leinhardt2012}. We define the collisions that are not in the hit-and-run regime and have $Q_{\rm R}>Q^{\prime*}_{\rm RD}$ as catastrophic collisions, which potentially cause the catastrophic disruption of the impactors~\citep{Stewart2009}. The collisions with $Q_{\rm R}>1.8Q^{\prime*}_{\rm RD}$, which corresponds to $M_{\rm lr}<0.1(M_{\rm l}+M_{\rm s})$ in Eq. (\ref{eq:3}), are super-catastrophic collisions~\citep{Leinhardt2012}.
 
The hit-and-run collisions cannot disrupt the larger impactor, but can transform the smaller impactor~\citep{Gabriel2023}. For collisions in the hit-and-run regime, the universal law for reverse collisions~\citep{Leinhardt2012} is used:
\begin{linenomath}
    \begin{equation}
        \label{eq:4}
        \frac{M_{\rm slr}}{M_{\rm s}+M_{\rm interact}}=-0.5\left(\frac{Q^\dagger_{\rm R}}{Q^{\dagger*}_{\rm RD}}-1\right)+0.5,
    \end{equation}
\end{linenomath}
where $M_{\rm slr}$ is the mass of the second largest remnant, which is the remnant of the smaller impactor, $M_{\rm interact}$ is the interacting mass from the larger impactor, which can be calculated using the masses of impactors and the geometry of the collision (impactor radii and impact angle)~\citep{Leinhardt2012}. $Q^\dagger_{\rm R}$ and $Q^{\dagger*}_{\rm RD}$ are the reduced mass kinetic energy and the critical disruption energy for reverse collisions calculated using $M_{\rm s}$, $M_{\rm interact}$, $\theta$, $c^*$, and $\bar\mu$. For hit-and-run collisions, we define the collisions that have $Q^\dagger_{\rm R}>Q^{\dagger*}_{\rm RD}$ as catastrophic hit-and-run collisions, which potentially disrupt the smaller impactor. The collisions with $Q^\dagger_{\rm R}>1.8Q^{\dagger*}_{\rm RD}$, which corresponds to $M_{\rm slr}<0.1(M_{\rm s}+M_{\rm interact})$ in Eq. (\ref{eq:4}), are super-catastrophic hit-and-run collisions~\citep{Leinhardt2012}.

Experiments of shock vaporization of iron indicate that vaporization of metal cores in a planetesimal collision requires an impact velocity higher than 18$\pm$5~km\,s$^{-1}$~\citep{Kraus2015}. This threshold of core vaporization would be lower in the ambient pressure of solar nebula~\citep{Davies2020}. For CB chondrites, we assume that they originated from super-catastrophic disruption events capable of completely fragmenting their pre-collision parent body. Therefore, super-catastrophic collisions and super-catastrophic hit-and-run collisions with impact velocities exceeding 13~km\,s$^{-1}$ are selected, which can cause vaporization of planetesimal cores and potentially formed the CB chondrules and metal grains. These collisions satisfy both criteria: an impact velocity exceeding 13~km\,s$^{-1}$ and a super-catastrophic outcome defined using the planetesimal masses. If a higher critical velocity such as 18~km\,s$^{-1}$ is assumed, fewer but not zero collisions exceed this threshold in most of our simulations. Given the rarity of CB chondrites, our results remain consistent with their formation. 

All collisions in our simulations are treated as perfect merging, although we have classified the potential outcomes of the collisions during the post-processing stage. The conditions of the collisions (impact velocities, impact angles, and identity of impactors), which are determined by the orbital evolution of planetesimals, have been accurately simulated. The perfect-merging simplification in the simulations might result in unrealistic size distribution of post-collision planetesimals because collisional fragmentation of planetesimals may produce a large number of small fragments~\citep{Durda2007,Sevecek2017} and planetesimals do not merge in hit-and-run collisions~\citep{Asphaug2010}. Nevertheless, our test simulations using different sizes of planetesimals and those with much larger initial planetary embryos yield similar results (Supplementary information section 1). These tests suggest that planetesimal size has limited effect on the high-velocity collisions and their driving forces. Therefore, neglecting fragmentation in collisions may not substantially alter our main conclusions. However, the perfect-merging assumption may result in shortened accretion timescale of planetary embryos due to neglecting hit-and-run collisions~\citep{Burger2020}. The Episode III collisions driven by grown planetary embryos potentially occur later if a realistic collision model is used.

\subsection{Calculation of secular resonance locations}\label{sec3-3}
The secular resonances of Jupiter and Saturn ($\nu_5$ and $\nu_6$) occur when the apsidal precession rate of a planetesimal ($g_{\rm p}$) matches that of Jupiter ($g_{\rm J}$) or Saturn ($g_{\rm S}$). In our simulations, both planetesimals and giant planets feel the gravitational potential of the gas disk. Therefore, $g_p$ includes the contributions of perturbation by Jupiter ($g_{\rm p,J}$), Saturn ($g_{\rm p,S}$), and the gas disk ($g_{\rm p,disk}$)~\citep{Nagasawa2005}:
\begin{linenomath}
    \begin{equation}
        \label{eq:5}
        g_{\rm p}=g_{\rm p,J}+g_{\rm p,S}+g_{\rm p,disk}.
    \end{equation}
\end{linenomath}
The perturbation by gas disk also contributes $g_{\rm J}$ and $g_{\rm S}$:
\begin{align}
g_{\rm J}=g_{\rm J,S}+g_{\rm J,disk}, \label{eq:6} \\
g_{\rm S}=g_{\rm S,J}+g_{\rm S,disk}, \label{eq:7} 
\end{align}
where $g_{\rm J,S}$ and $g_{\rm J,disk}$ are the precession rates of Jupiter due to Saturn and the gas disk, and $g_{\rm S,J}$ and $g_{\rm S,disk}$ are the precession rates of Saturn due to Jupiter and the gas disk.
 
We calculated these precession rates following the Appendix of ref.~\citep{Nagasawa2000}. The precession rates due to the perturbation by the gas disk ($g_{\rm p,disk}$, $g_{\rm S,disk}$, and $g_{\rm J,disk}$) were calculated by numerically integrating the disturbing function $T(a,t)$ over the gas disk in three dimensions. $T(a,t)$ varies with semi-major axis ($a$) and time ($t$). Alternatively, $g_{\rm p,disk}$, $g_{\rm S,disk}$, and $g_{\rm J,disk}$ can be calculated analytically~\citep{Ward1981, Nagasawa2005}:
\begin{linenomath}
    \begin{equation}
        \label{eq:8}
        g_{\rm disk}\sim-Z_k\left[\frac{\pi G\Sigma_{\rm gas}(a,t)}{na}\right],
    \end{equation}
\end{linenomath}
where $Z_k=1$ (for the exponent of $-1$ in Eq. \eqref{eq:2}), $G$ is the gravitational constant, $n$ is the mean motion of the planetesimal, Jupiter, or Saturn. The locations of $\nu_5$ and $\nu_6$ calculated by numerically integrating $T(a,t)$ are consistent with those calculated using Eq. (\ref{eq:8}) (Supplementary Fig.~12). Note that both methods assume small eccentricities for planetesimals and neglect planetesimal-planetesimal interactions, thus the results are the first-order estimates of the locations of $\nu_5$ and $\nu_6$ in our simulations.

\backmatter

\clearpage
\begin{appendices}

\section{Data Availability}
The collision data generated in this study have been deposited in the publicly available
repository: https://doi.org/10.5281/zenodo.19970387. Source data are provided with this paper.

\section{Code Availability}
N-body simulations were performed using GENGA, publicly available at https://bitbucket.org/sigrimm/genga.

\clearpage
\section{Acknowledgements}
The support and resources from the High-Performance Computing Center (HPCC) of Nanjing University and High-Performance Computing Center of Collaborative Innovation Center of Advanced Microstructures are gratefully acknowledged.

\section{Funding}
This work was supported by National Natural Science Foundation of China (NSFC; Nos. 12373081 and 12150009 to L.-Y.Z., No. 42508025 to K.S.), the China Postdoctoral Science Foundation (CPSF; No. 2025M773192 to K.S.), the Postdoctoral Fellowship Program of CPSF (No. GZC20252094 to K.S.), and Jiangsu Funding Program for Excellent Postdoctoral Talent (No. 2025ZB246 to K.S.).

\section{Author contributions}
K.S. and L.-Y.Z. conceived the study. K.S. performed N-body simulations. L.-Y.Z. and H.H. provided computing resources. K.S. and L.-Y.Z. contributed formal analysis. All authors contributed to the discussion of physical implications and manuscript preparation.

\section{Competing interests}
The authors declare no competing interests.

\clearpage
\section{Supplementary information}\label{secA2}
\subsection{Supplementary information}
Supplementary Methods and Figs. 1--12. Test simulations for different initial conditions (resolutions, planetesimal sizes, inclusion of initial embryos, and initial giant planet orbits) and model configurations (gas disk depletion models, neglect‌ing gas disk potential and planetesimal-planetesimal interactions).

\subsection{Supplementary Movie \rvideo{v1}}
Orbits of projectiles in catastrophic collisions. The instantaneous orbits of projectiles in catastrophic collisions that occur between 1.8 and 2.2~Myr and have impact velocities >13~km\,s$^{-1}$ are shown. The dots represent the collision locations. The orbits and collisions are color-coded as in Fig.~\ref{figure2}. The projectiles are first locked in $\nu_5$, with the same apsidal precession rate as that of Jupiter. Obtaining large eccentricities, these projectiles escape from $\nu_5$ successively. Eventually, they migrate inward and collide with other planetesimals.

\subsection{Supplementary Movie \rvideo{v2}}
Evolution of semi-major axes and eccentricities for all planetesimals in Episode I. The results of the simulation with $\tau=1$~Myr are shown. The video starts at 1~Myr and ends at 2.4~Myr. The projectiles in catastrophic collisions are color-coded as in Fig.~\ref{figure2}.

\subsection{Supplementary Movie \rvideo{v3}}
Evolution of the critical angle of secular resonance $\nu_5$ for projectiles in Episode I catastrophic collisions. The results of the simulation with $\tau=1$~Myr are shown. The video starts at 0.5~Myr and ends at 2.4~Myr. The projectiles are color-coded as in Fig.~\ref{figure2}.

\end{appendices}

\end{document}